\documentclass[12pt,a4paper]{article} 
\usepackage{epsfig} 
\usepackage{amsmath}
  \usepackage{cite}

\def\nostrocostrutto#1\over#2{\mathrel{\mathop{\kern 0pt \rlap  
  {\raise.2ex\hbox{$#1$}}} 
  \lower.9ex\hbox{\kern-.190em $#2$}}} 
 
\def\epem{e^+e^-} 
\def\bb{b\bar{b}} 
\def\nhad{{N}^{ch}_{had}} 
\def\nb{{N}^{ch}_{b}} 
\def\nc{{N}^{ch}_{c}} 
\def\nl{{N}^{ch}_{\ell}} 
\def\cc{\rm c\bar{c}} 
\def\light{\rm u\bar{u},d\bar{d},s\bar{s}} 
\def\dcl{\delta_{c\ell}} 
\def\dbl{\delta_{b\ell}} 
 
\def\cN{{\cal{N}}} 
\def\cF{{\cal{F}}} 
 
\begin{document} 
 
\setcounter{page}{0} 
\thispagestyle{empty} 
 
\def\jet{{\mbox{\scriptsize jet}}} 
\def\cO#1{{\cal O}\left(#1\right)} 
\def\lrang#1{\left\langle#1\right\rangle} 
\def\br{brems\-strah\-lung} 
\def\Br{Brems\-strah\-lung} 
\def\QQ{\relax\ifmmode{Q\overline{Q}}\else{$Q\overline{Q}${ }}\fi} 
\def\qq{\relax\ifmmode{q\bar q}\else{$q\bar q${ }}\fi} 
\def\abs#1{\left|#1\right|} 
 
\def\la{\mathrel{\mathpalette\fun <}} 
\def\ga{\mathrel{\mathpalette\fun >}} 
\def\fun#1#2{\lower3.6pt\vbox{\baselineskip0pt\lineskip.9pt 
  \ialign{$\mathsurround=0pt#1\hfil##\hfil$\crcr#2\crcr\sim\crcr}}} 
 
\def\eV{{\rm e\kern-0.12em V}} 
\def\MeV{{\rm M}\eV} \def\keV{{\rm k}\eV} 
\def\GeV{{\rm G}\eV} \def\TeV{{\rm T}\eV}

\newcommand\beeq{\begin{eqnarray}} 
\newcommand\eeeq{\end{eqnarray}} 
\def\as{\alpha_s}

\title{\bf Multiplicity Difference between Heavy and Light 
Quark Jets Revisited} 
 
\vspace{7cm} 
 
\author{Yuri L. Dokshitzer$^{1,2}$, Fabrizio Fabbri$^3$, %
\ Valery A. Khoze$^{4,2}$ 
\ and \\ 
 \ Wolfgang Ochs$^5$ 
} 
  
\date{ 
{\normalsize\it  
$^{1)}$LPTHE, University Paris-VI, 4 place Jussieu, F-75252 Paris, France\\ 
$^{2)}$ PNPI, Gatchina, St. Petersburg, 188300, Russia\\ 
$^{3)}$ INFN e Dipartimento di Fisica dell'Universit\`a di Bologna,\\ 
v.le Berti Pichat 6/2, 40127 Bologna, Italy \\ 
$^{4)}$ Department of Physics and Institute for Particle Physics 
Phenomenology\\ 
University of Durham, Durham, DH1 3LE, UK \\  
$^{5)} $Max-Planck-Institut f\"ur Physik (Werner-Heisenberg-Institut)\\ 
F\"ohringer Ring 6, D-80805 M\"unchen, Germany\\  
\vspace{1.0cm} 
}} 
 
\maketitle 
 
\thispagestyle{empty} 
\begin{abstract} 
The perturbative QCD approach to multiparticle production predicts a 
characteristic suppression of particle multiplicity in a heavy quark jet  
as compared to a light quark jet. In the Modified Leading Logarithmic 
Approximation (MLLA) the multiplicity difference $\delta_{Q\ell}$ between 
heavy and light quark jets is derived in terms of a few other  
experimentally measured 
quantities. The earlier prediction for $b$-quarks  needs revision in the 
light of new experimental results and the improvement in the understanding of 
the experimental data. We now find $\delta_{b\ell}=4.4\pm0.4$.  
The updated MLLA results on  $\delta_{b\ell}$ and  $\delta_{c\ell}$ 
are compared with the present data from  
$e^+e^-$ annihilation. 
Their expected energy independence is confirmed within the energy range 
between 29 and 200 GeV; the absolute values are now in better agreement  
with experiment  
than in the previous analysis, and the remaining difference can be  
attributed largely to next-to-MLLA contributions, an important subset of
which are identified and evaluated. 
\end{abstract} 
 
\vspace{-22cm}
 
\rightline{DFUB 2005-09; DCPT/05/92; IPPP/05/46}
\rightline{LPTHE-05-22; MPP-2005-61}
\vspace{-20cm}

\newpage 
 
\section{Introduction} 
 
 Since the early days of QCD, heavy quark physics has been one of the 
 primary testing grounds for many aspects of the theory. In the last 
 years a wealth of new important results on the profile of jets 
 initiated by heavy quarks $Q(b,c)$ has been reported  by the 
 experimental collaborations at LEP, SLC, Tevatron and HERA. Future 
 progress is expected from the measurements at the LHC and a future 
 linear $\epem$ collider.  
These studies are important for the tests of  
the basic concepts of the QCD description  
of multiparticle production and also for the studies of new physics. 
 
Multiple hadron production in hard processes is derived from the QCD 
parton cascade processes which are dominated by gluon bremsstrahlung.  
 An essential difference in the structure of the energetic heavy and 
 light quark jets $(\ell\equiv q=u,d,s)$ results from the  
  dynamical 
 restriction on the phase space of primary gluon radiation in the 
 heavy quark case:  
 the gluon radiation off an energetic quark $Q$ with mass $M$ and energy 
 $E_Q\gg M$ 
 is suppressed inside the forward angular cone with an opening angle 
 $\Theta_0 = M/E_Q$,  
the so-called dead cone phenomenon \cite{dkt1,dkt7}.  
This is in close analogy with QED where the photon radiation is suppressed at 
small angles with respect to a primary charged massive particle. 
The suppression of 
 the energetic gluon emission at low momentum transfer $k_\bot$  
 results, in turn, in the 
 decrease of the heavy quark energy losses. This provides a 
 pQCD 
 explanation of the leading particle effect~\cite{dktHQ,CG} which is 
 clearly seen experimentally in the $b\bar b$ and $c\bar c$ events in 
 $\epem$ annihilation~\cite{pdg}; for recent reviews, see \cite{ko}. 
 
For a long time, there has been no clear explicit experimental visualisation 
 of the dead cone. Only recently, preliminary DELPHI results have been 
reported~\cite{vpn} which show the expected depletion of small angle 
particle production in $b$-jets with respect to the heavy hadron direction.  
where the decay vertices of the heavy hadrons were reconstructed. 
Further detailed studies of the dead cone effect in different processes 
are needed. 
Some new results may come from the current analysis  
of the structure of the $c$-quark jets,  
produced in the photon gluon fusion in Deep Inelastic Scattering at HERA.  
 
 It is worthwhile to mention that the difference in the radiation from 
 massive and massless quarks   
 should 
 also 
 manifest itself in the QCD medium  
  via suppression of the 
 {\em medium-induced  
 radiative 
 energy loss}\/ of heavy quarks propagating in a strongly interacting 
 matter, see, for example, \cite{dkh,tks,asw,wcx} and references therein.

Studies of heavy quark jets are also important in the investigation 
of the properties 
of known or new heavy objects. For example, 
a detailed knowledge of the $b$-jet profile is needed 
 for the analysis of the final state in the $t\bar t$ production 
 processes. Various aspects of studying new physics, in particular, of 
 the structure of the Higgs sector at the LHC and at a future linear collider, 
 would benefit from the detailed understanding of the $b$-initiated jets, 
see for example \cite{ksw,bkso}.

 The dead cone phenomenon leads to  
 essential differences in the profiles of the light- and 
 heavy-quark-initiated jets. According to the concept of ``Local Parton Hadron 
Duality'' (LPHD)~\cite{adkt1},  
 the dead cone suppression of gluon radiation should result in the 
 characteristic differences in 'companion' 
 spectra and multiplicities 
 of primary light hadrons in these jets \cite{dkt1,dkt7,bas}. 
 
 In particular, 
 as a direct consequence of the LPHD scenario, the difference of 
 companion multiplicities 
  $N^{h}$ 
 of light hadrons in the heavy quark  
  and  
 light-quark jets at the same jet energy $E_\jet$ should be {\em 
 energy independent}\/ (up to a power correction $ 
 \cO{M^2/E^2_{\jet}}$), i.e. in $\epem$ annihilation at c.m.s. energy  
$W=2E_{\jet}$ one obtains the QCD prediction~\cite{bas,ko} 
\begin{equation} 
N_{q\bar q}^h(W)-N_{\QQ}^h(W)  
    \  = \ \text{const} (W). 
\label{Nconst} 
\end{equation} 
 The corresponding constant is different for $c$- and $b$-quarks and 
 depends on the type of light hadrons  
  $h$  
 under study.  
 
This 
 prediction 
 is in marked contrast with the  
 expectation 
 of the so-called na\"\i ve model \cite{pcr}, 
 which relates the multiplicities in light and heavy quark events  
%
based on the idea of the {\em reduction of the energy scale}, 
\begin{equation} 
 N^h_{\QQ}(W)=N^h_{q\bar q}\left((1\!-\!\langle x_Q\rangle)W\right); \>\> 
 \lrang{x_Q}\! = \!\frac{2\lrang{E_Q}}{W}, \>\> 1\!-\!\lrang{x_Q} = 
 \cO{\alpha_s(W)}.  
\end{equation} 
 In this case 
 the difference of $q$- and $Q$-induced multiplicities  
would  {\em grow}\/ gradually with $W$ as 
\begin{equation} 
N_{q\bar q}^h(W)-N_{\QQ}^h(W)  
    \  \propto\  \sqrt{\alpha_s}\ln\frac{1}{\alpha_s} \cdot N_{q\bar q}^h(W). 
\label{grow} 
\end{equation} 
 
 In this paper we focus on the analysis of the current experimental 
 status of the difference of the average charged multiplicities 
 $\delta_{b\ell}$ of events containing $b$- and light quarks in 
 $\epem$ annihilation in the available energy range. The situation with 
charmed quarks is considered as well. 
The main emphasis 
 is on the comparison between the reanalysed data and the expectations 
 based on the MLLA, in an extension of previous analyses \cite{bas,ko}.  
In addition, we discuss the size of 
the next-to-MLLA contributions. 
 
\section{Theoretical Analysis} 
 
Within the LPHD framework, the multiplicity of light hadrons in $\epem$ 
annihilation events is proportional to that of \br\ partons. 
To predict the QCD yield of light particles accompanying $\QQ$ 
production we have first to address the question on how the 
development of the parton cascade initiated by a heavy quark $Q$ depends 
on the quark mass $M$.

\subsection{Structure of QCD Cascades in $e^+e^- \to Q\overline{Q} + \ldots$} 
 
As well known, in the case of a light quark jet the structure of the 
parton branching of the primary gluon $g_1$ with energy $\omega_1$ 
(energy spectra, multiplicities of secondaries) is determined by the 
parameter 
\begin{equation} 
 \kappa_q=4 \omega^2_1\sin^2 \frac{\Theta_1}{2}\/, 
\label{eq1} 
\end{equation} 
 where $\Theta_1$ is the angle between the gluon and the energetic 
 quark.  
 (The expression \eqref{eq1} is 
written 
in such a way as to account for the next-to-leading correction due to 
large-angle soft gluon emission, up to the full jet opening angle 
$\Theta_1=\pi$, see for example~\cite{dkmt1,do}.) 
 For $\Theta_1\ll 1$ this parameter reduces to the gluon transverse 
 momentum, $k_t^2\simeq(\omega_1\Theta_1)^2$.  
The appearance of this scale 
is a consequence of colour coherence in multiplication of soft gluons 
which dominate the QCD cascades. This destructive coherence results in 
the {\em Angular Ordering}\/ (AO) of successive parton 
branchings~\cite{AO}.

The corresponding parameter for a jet initiated by a {\em heavy}\/ 
quark with energy $E_Q$ and the mass $M$ reads 
\begin{equation} 
 \kappa_Q\>=\> \omega^2_1 
 \left[\,\left(2\sin\frac{\Theta_1}{2}\right)^2 +\Theta_0^2\,\right]\/; 
\qquad \Theta_0 \equiv \frac{M}{E_Q}.  
\label{eq2} 
\end{equation} 
%
Note that the same quantity $\kappa_Q$ determines the scale of the 
running coupling in the gluon emission off the massive 
quark.\footnote{A detailed analysis of the running coupling argument 
  in the massive quark case can be found in Appendices 
  to~\cite{dktHQ}, see also~\cite{dkmt2,ko}.} 
 

The modification of the angular parameter in (\ref{eq2}) caused by the 
heavy quark mass has a transparent physical 
interpretation.\footnote{This argument is based on the discussion of 
  two of the authors (YLD,VAK) with S.I.~Troyan in the early 90's. 
}   
Consider radiation of a secondary gluon $g_2$ with energy 
$\omega_2\ll\omega_1$ at angle $\Theta_{21}$ relative to the primary 
gluon $g_1$. 
 Normally, in the ``disordered'' angular kinematics, 
 $\Theta_{21}>\Theta_{1}$, the destructive interference between 
 the emission amplitudes of $g_2$ off the quark and $g_1$ {\em 
 cancels}\/ the independent radiation $g_1\!\to\!g_2$ thus enforcing 
\begin{equation} 
 \Theta_{21}\le\Theta_{1}\,. 
\label{eq:AO} 
\end{equation} 
 Meantime, in the massive quark case the interference contribution 
 enters the game only when the angle $\Theta_2$ of $g_2$ with respect 
 to the {\em quark}\/ is larger than the dead cone, 
 $\Theta_{2}>\Theta_0$. Therefore, the cancellation leading to the AO 
 condition \eqref{eq:AO} does not occur when the gluon $g_1$ is 
 radiated {\em inside}\/ the dead cone, $\Theta_1<\Theta_0$, and the 
 jet evolution parameter \eqref{eq2} {\em freezes}\/ in the 
 $\Theta_1\to0$ limit.


 In physical terms what happens is the {\em loss of coherence}\/ 
 between $Q$ and $g_1$ as emitters of the soft gluon $g_2$ due to 
 accumulated longitudinal separation $\Delta z 
 >\lambda_{||}^{(2)}\approx \omega_2^{-1}$ between the massive and 
 massless charges ($v_Q\approx 1-\Theta_0^2/2 <1$, $v_1=1$).  Indeed, 
 during the formation time of the secondary radiation, 
 $t_{f}^{(2)}\sim (\omega_2\Theta_{21}^{2})^{-1}$, the two sources --- 
 the quark and the gluon $g_1$ --- separate in the longitudinal 
 direction by 
\begin{equation}\label{tformarg} 
 \Delta z 
 \>\sim\> t_{f}^{(2)} \abs{v_Q - c\cos\Theta_{1}} \>\simeq\> 
 \lambda_{||}^{(2)} \cdot 
 \frac{\Theta_{1}^2+\Theta_0^2}{\Theta_{21}^2}\,. 
\end{equation} 
 It is the last factor that determines whether an interference is 
 essential or not. When this ratio is larger than 1, the quark $Q$ and 
 gluon $g_1$ are separated enough for $g_2$ to be able to resolve the 
 two emitters as independent colour charges. In these circumstances 
 $g_1$ acts as an independent source of the next generation \br\ 
 quanta. Otherwise, no additional particles triggered by $g_1$ emerge 
 on top of the yield determined by the {\em quark}\/ charge (which 
 equals the total colour charge of the $Q+g_1$ system). 
 
 In the massless quark case ($\Theta_0\equiv0$) this consideration 
 reproduces the standard AO prescription \eqref{eq:AO}. 
 In the massive quark case, the separation ({\em incoherence}\/) 
 condition $\Theta_{21}^2\le (\Theta_1^2+\Theta_0^2)$ results in 
 \eqref{eq2} as the proper evolution parameter for the gluon subjet. 
 
 The modification \eqref{eq2} may look superfluous since the soft 
 gluon radiation inside the dead cone, $\Theta_1\ll\Theta_0$, is 
 suppressed. In spite of this, it is essential for keeping track of 
 the next-to-leading order (MLLA) corrections in accompanying 
 multiplicities. 
 In the Appendix A we recall the structure of the exact matrix 
 element for gluon radiation off a heavy \QQ pair and show how the 
 parameter \eqref{eq2} naturally appears in the problem.

 \subsection{MLLA prediction for accompanying multiplicity and its 
 accuracy} 
 
 The light charged hadron multiplicity in heavy quark $\epem$ 
 annihilation events at c.m.s. energy $W$ can be represented as 
\begin{equation} 
N_Q^{ch}(W)\equiv  N_{\epem \to \QQ}^{ch}\, (W) \>=\> N_{\QQ}^{ch}\, (W)+n^{dc}_Q, 
\label{eq4} 
\end{equation} 
where $N_Q^{ch}$ is the charged multiplicity of $e^+e^-$ events containing 
a heavy quark~$Q$; 
 $N_{\QQ}^{ch}\,(W)$ is the charged multiplicity of light hadrons 
accompanying the heavy quark production 
 (excluding decay products of $Q$-flavoured hadrons) and $n^{dc}_Q$ 
 stands for the constant charged decay multiplicity of the two leading heavy 
 hadrons ($n^{dc}_b=11.0\pm0.2$ for $b$-quarks, $n_c^{dc} =5.2\pm 0.3$ 
 for $c$-quarks, see for example~\cite{bas} for details of a previous
evaluation). 
 As shown in Appendix A, at $W=2E_Q\gg M\gg\Lambda_{\mbox{\scriptsize 
     QCD}}$ the companion multiplicity $ N_{\QQ}(W)$  
can be related to the particle yield in the 
 {\em light quark}\/ events $\epem\to q\bar q$ ($q=u,d,s$) as 
 \cite{bas,ko} 
\begin{equation} 
 N_{q\bar q}(W) \>\,-\,  N_{\QQ}(W)\> =\>  N_{q\bar q} 
 (\sqrt{e}M)\cdot \left[\,1+\cO{\alpha_s(M)}\,\right], 
\label{eq3} 
\end{equation} 
where we approximately expressed the {\em difference}\/ between 
the light- and heavy-quark generated multiplicities in terms of the 
light-quark event multiplicity at reduced ($W$ independent) c.m.s. 
energy $W_0=\sqrt{e}M$, $e=\exp(1)$. 
  
 Concerning the {\em accuracy}\/ of \eqref{eq3}, there are two 
 separate issues one has to address, namely: 
\begin{enumerate} 
\item the accuracy of the statement of the {\em constancy}\/ of 
 the l.h.s.\ of \eqref{eq3},  
\item to which accuracy this difference can be quantitatively 
  predicted by means of pQCD (the r.h.s.). 
\end{enumerate}

\paragraph{Left-Hand-Side.} 
 Answering the first question, it turns out to be insufficient  
to compare particle multiplicities in a 
 given order of perturbation theory. 
 Indeed, within the next-to-leading accuracy (MLLA), for example, one 
 takes into consideration (``exponentiated'') $\sqrt{\alpha_s 
 }+\alpha_s$ effects in the anomalous dimension describing parton 
 cascading, and $1+\sqrt{\alpha_s}$ terms in the normalisation 
 (coefficient functions). This allows to predict the l.h.s.\ of 
 \eqref{eq3} 
 up to the NNLO correction the absolute magnitude of which is 
of the order of 
\begin{equation}\label{eq:cnstMLLA} 
  (\mbox{l.h.s.}) - (\mbox{l.h.s.})^{MLLA} \>=\> \cO{\alpha_s(W)\cdot 
  N_{q\bar{q}}(W)}. 
\end{equation} 
Meantime, the steep growth with energy of the multiplicity factor 
$N(W)$ (faster than any power of $\ln W$) makes the neglected 
$\alpha_s N(W)$ correction {\em dominate}\/ over the (presumably) 
finite r.h.s.\ in \eqref{eq3}, thus endangering the very possibility 
of discriminating between $Q$- and $q$-jet multiplicities. 
 
However, examining the origin of perturbative corrections proportional 
to $N(W)$ in \eqref{eq:cnstMLLA} one can see that all of them prove to 
be independent of the quark mass $M$, being inherent to the light 
quark jet evolution itself. For example, the first corrections of the 
order of $\alpha_s(W)N(W)$ to the MLLA expression \eqref{eq:cnstMLLA} 
come either from further improvement of the description of the 
anomalous dimension $\Delta\gamma(\alpha_s)\sim\alpha_s^{3/2}$  
determining 
intrajet cascades, or from $O(\alpha_s(W))$ terms in the coefficient 
function due to 
\begin{itemize} 
 \item three-jet configuration {\em quark } + {\em antiquark } + {\em 
 hard gluon at large angle},  
 \item the so-called ``dipole correction'' to the AO scheme {\em quark 
 } + {\em antiquark } + {\em two soft gluons at large emission angles 
 } (see \cite{dkmt2}), 
\end{itemize} 
 both of which are insensitive to the $\Theta_0$ value with {\em power 
 }\/ accuracy $\cO{\Theta_0^2}\!\ll\!1$. 
 
 In fact, the statement that the l.h.s.\ of \eqref{eq3} does not 
 depend on the annihilation energy follows from general considerations and 
 should hold {\em in all orders}\/ in perturbation theory with power 
 accuracy, $1+\cO{M^2/W^2}$. 
 
 This is a very powerful statement which goes beyond the standard 
 renormalisation group (RG) wisdom about separation of two 
 parametrically different scales, $W$ and $M$. Indeed, by looking upon 
 the particle multiplicity as a moment ($N=0$) of the inclusive 
 fragmentation function, and by drawing an analogy with the OPE 
 analysis of DIS structure functions (space-like parton 
 distributions), one could expect for light- and 
 heavy-quark initiated multiplicities 
\begin{equation} 
 \frac{N_{q\bar q}(W)}{N_{\QQ}(W)} \>=\> f(M) = \mbox{const}(W)\,, 
\label{eq4wr} 
\end{equation} 
 that is that their {\em ratio}\/ rather than the {\em difference}\/ 
 is $W$-independent. The RG motivated expectation \eqref{eq4wr} would 
 have been correct if the quark mass $M$ played the r\^ole of the 
 initial condition for parton evolution --- the {\em transverse 
 momentum cut-off}. This is true enough for {\em hard}\/ gluons with 
 energies $x=2\omega/W\sim 1$ for which the region $k_\perp < M$ is 
 indeed suppressed as compared to the massless quark case. 
 It is not hard gluons that dominate the accompanying multiplicity 
 however.  
 
 Turning to (primary) gluons with $x\ll 1$ we observe that the 
 radiation off light and heavy quarks {\em remains the same}\/ down to 
 much smaller transverse momentum scales namely, 
$$ 
 k_\perp \>\ga\> \omega\cdot \frac{2M}{W} = xM\>\ll \> M\,, 
$$ 
 which is nothing but the statement of the ``dead cone'' suppression 
 discussed above. 
 It is important to stress that, being based on the analysis of the 
 first order gluon radiation matrix element, this conclusion is {\em 
 exact}\/ and holds {\em in all orders}\/ in perturbative 
 expansion. This follows from the fact that emission of gluons with 
 $x\ll 1$ is governed by the Low--Burnett--Kroll theorem~\cite{LBK} concerning the 
 classical nature of soft accompanying radiation (following the 
 $dx\cdot(1/x-1)$ distribution), which holds to power 
 accuracy, see also \cite{dks}. 
 
 So  
 the QCD coherence plays a fundamental r\^ole in establishing this 
 result~\cite{dfk1}.  Since the gluon \br\ off {\em massive }\/ and 
 {\em massless }\/ quarks differs only at {\em parametrically small 
   angles}\/ $\Theta\la\Theta_0 $, the AO (QCD coherence) then ensures 
 that the accompanying cascading effects 
 are limited from above by a finite 
 factor $N(W\cdot\Theta_0)\simeq N(M)$. 
%
 A rigorous proof of the statement that $W$ dependent corrections to 
 the r.h.s\ of \eqref{eq3} are power suppressed as $M^2/W^2$ (of 
 subleading twist nature, in the OPE nomenclature) 
 is lacking at the moment. 
 
By replacing the approximate MLLA multiplicities in \eqref{eq:cnstMLLA} 
by the experimentally  observable multiplicities in \eqref{eq3} it becomes possible 
to establish a phenomenological relation between the light and heavy quark 
jets with controllable accuracy. 
 
Thus, the difference in the mean charged multiplicities, 
$\delta_{Q\ell}$, between heavy and light quark events at fixed 
annihilation energy $W$ depends only on the heavy quark mass $M$ and 
remains $W$-independent (with power accuracy)~\cite{dfk1,bas} 
\begin{eqnarray} 
\delta_{Q\ell} &=\  N^{ch}_{Q}\ (W)  
 - N^{ch}_{q}\ (W)  &= \ {\rm const}\ (W), 
 \label{delql}\\ 
\delta_{bc} &=\  N^{ch}_{b}\ (W) 
 - N^{ch}_{c}\ (W) &=\ {\rm const}\ (W) ,  
\label{delbc} 
\end{eqnarray} 
with $Q=b,c$ and $\ell\equiv q=u,d,s$.

\paragraph{Right-Hand-Side.} 
 
 The r.h.s.\ of \eqref{eq3} is estimated with the MLLA accuracy. In 
 general, this constant difference is proportional to $N(M)$ and can 
 be given in terms of the series in $\sqrt{\alpha_s(M)}$ as the pQCD 
 expansion parameter. Let us remark that such an expansion formally 
 relies upon treating the quark mass $M$ as the second hard scale, 
 $\alpha_s(M)\ll1$, and is bound to be only moderately satisfactory at 
 best, since in practice, in \eqref{eq3}, the bottom quark mass translates  
into $W_0^b\sim 8\,\GeV$ 
and the charm quark mass into $W_0^c\sim 2.5$ GeV only.%
\footnote{Short-lived top quarks do not follow this pattern 
in the first place; the $N_{t\bar{t}}$ notion being 
 elusive, see for example \cite{bigi,kos,dkt1}.}

 \subsection{Quark mass effects in three-jet events} 
\label{quarkmass} 
 Another powerful, and phenomenologically interesting, consequence of 
 QCD coherence is that the structure of particle cascades in three-jet 
 $\QQ g$ events (with a hard gluon radiated at large angle) must 
 be identical to that in the light-quark case everywhere, apart from 
 the two narrow angular regions corresponding to the dead cones of the 
 $Q$-quarks. More specifically, the particle multiplicity in 3-jet 
events can be written in MLLA as the sum of quark and gluon jet multiplicities 
\cite{dkt88} (see also \cite{egk}) 
\begin{equation} 
N_{q\bar q g}(W)= N_{q\bar q}(2E_q^*)+\frac{1}{2} N_{gg}(p_\perp^*),\label{Nqg} 
\end{equation} 
where $E_q^*$ denotes the $q$ or $\bar q$ energy 
and $p_\perp^*$ the gluon transverse momentum, both in the  c.m.s. frame 
of the $\qq$ pair. Then, with $W_{\qq}=2E^*_q$, we obtain 
\begin{equation} 
N_{\QQ g}(W) - N_{q\bar q g}(W)=N_{\QQ}(W_\QQ) - 
    N_{q\bar q}(W_{q\bar q}). \label{Qqg} 
\end{equation} 
 This may provide another handle for the 
 detailed studies of the dead cone phenomenon at the reduced effective c.m.s.\ 
 energies $W_{\QQ}$.

 \subsection{Discussion and estimate of next-to-MLLA 
 terms of the order of ${\alpha_s(M)N(M)}$ } 
 
 In \cite{vap} the equation \eqref{eq3} was evaluated exactly which
constituted an attempt to improve the pQCD prediction beyond the
 $\sqrt{\alpha_s}$ accuracy beyond which  \eqref{eq3} does not actually
hold.
 However, the main assumption of 
 \cite{vap} that the companion multiplicity is generated by a single 
 cascading gluon is not valid at this  level. 
 
 Next-to-MLLA correction terms are copious and it is hard to 
 collect them all. There are, however, some specific contributions that look 
 {\em enhanced}\/ and can be listed and estimated. These are 
 contributions containing an additional (semi-dimensional) factor 
 $\pi^2$. 
  
 In particular, to predict the event multiplicity at the $\alpha_s N$ 
 level, one has to take into consideration large angle two soft gluon 
 systems (aforementioned dipole configurations). This problem is 
 discussed in Appendix~A where we show how a $\pi^2$ enhanced 
 correction emerges.   
 Another correction of similar nature comes from the $1-z$ 
 rescaling of the argument of the dead cone subtraction. This 
 contribution is also extracted and analysed in Appendix~A. It turns 
 out to be numerically larger than the ``dipole'' contribution. 
 These enhanced next-to-MLLA effects work in the same direction: 
 they all tend to {\em increase}\/ the difference between the light and heavy 
 quark-initiated multiplicities in \eqref{eq3}.


\section{Theoretical predictions confronted with experiment} 
\subsection{Experimental results on heavy quark multiplicities 
in $\epem$ annihilation} 
 
The experimental measurements of hadron multiplicities in $b\bar b$  
and $c\bar c$ events produced in $\epem$ annihilation were performed in  
the wide range of c.m.s. energies $\sqrt{s} \equiv W$ from PEP, at  
$\sqrt{s} = 29$~GeV, to LEP2 at $\sqrt{s} = 206$~GeV 
\cite{row,delco,hrs,tpc,tasso1,tasso2,topaz,venus,mark2,opal1,sld1,delphi1,opal2,sld2,delphi2,opal3,sld3}. 
For reviews on this topic see, for instance,
\cite{chrin,ada1,ada2}. 
Within the experimental uncertainties the data clearly show that the differences 
$\delta_{b\ell}$ and $\delta_{c\ell}$  
are fairly independent of the c.m.s. energy, as expected from the perturbative 
analysis,  
and in a marked contrast to the steeply rising total mean multiplicity $N^{ch}_{had}$. 
 
\begin{figure}[t!]
\begin{center} 
\mbox{\epsfig{file=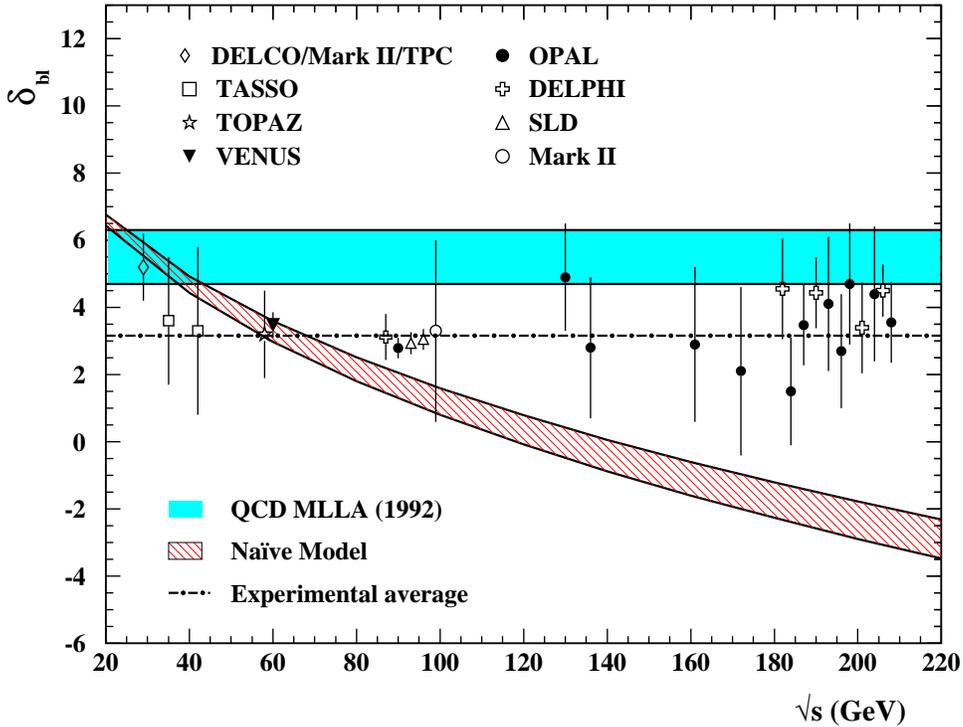,bbllx=1.0cm,bblly=7.cm,bburx=20.5cm,%
bbury=22.cm,width=13.0cm}} 
\end{center} 
\vspace{-0.3cm} 
\caption{ 
Experimental measurements of $\dbl$ plotted as a function 
of the c.m.s. energy~
\cite{row,delco,hrs,tpc,tasso1,tasso2,topaz,venus,mark2,
opal1,sld1,delphi1,opal2,sld2,delphi2,opal3,sld3}.
The 1992 MLLA expectation  
$\delta_{b\ell}^{MLLA}=5.5\pm0.8$ \cite{bas} (shaded area) includes experimental errors on  
$n_b^{dc}$ and light quark multiplicities at $\protect\sqrt{s}\simeq 8$ GeV.   
The prediction of the ``na\"\i ve model'' \protect\cite{pcr}  
based on the reduction of the energy scale is also shown (dashed area).} 
\label{fig1}  
\end{figure} 
 
This can be seen for example in Fig. 1, which shows a  
compilation of direct measurements of $\delta_{b\ell}$, Eq. (\ref{delql}). 
This figure is taken from \cite{opal3} with the addition of the result from the VENUS 
experiment at $\sqrt{s} = 58$~GeV \cite{venus} as well as the preliminary result from  
DELPHI at $\sqrt{s} = 206$~GeV \cite{delprel} . 
The dash-dotted line shown in Fig. 1 corresponds to the weighted average among all  
published results, $\delta_{b\ell}^{exp} = 3.12 \pm 0.14$, assuming that the  
measurements are uncorrelated. 
 
%
%
%

It is worthwhile to mention that the first preliminary data on the  
multiplicity difference between the $b$-quark and $udsc$-quark large-angle  
3-jet events produced in $Z^0$ decays, are reported by DELPHI \cite{ms,shd}. 
According to (\ref{Qqg}) these results can be related to the multiplicity difference, 
$\delta_{bq'}$, between the $b$-quark and the $q'$-quark ($q'=u,d,s,c$) events in 
$\epem$ annihilation measured in the effective energy range $W_{\QQ}\sim 53 - 59$ GeV. 
The data points do not show any sizeable energy dependence and are consistent
with the precise direct result from the VENUS experiment~\cite{venus} at $\sqrt{s}= 58$~GeV.  

 
As it can be seen in Fig. 1, within the experimental uncertainties most data points 
are consistent with the original MLLA prediction~\cite{bas},  
$\delta_{b\ell}^{MLLA} = 5.5 \pm 0.8$. 
However, the precise results from the OPAL, SLD, DELPHI and VENUS 
experiments,  which dominate  
the weighted average value  $\delta_{b\ell}^{exp}$, are definitely lower.   
The ``na\"\i ve model'', 
based on the reduction of the energy scale  $\sqrt{s}$, 
which predicts the growing difference as in Eq. (\ref{grow}) and, 
therefore,  
the gradually decreasing $\delta_{b\ell}$ is strongly disfavoured. 
 
\subsection{Test of MLLA predictions for $b$ quark jets} 
Our main goal here is to explain why the previous  
numerical value of the MLLA prediction of 
$\delta_{b\ell}^{MLLA} = 5.5 \pm 0.8$ \cite{bas} needs a revision. 
This value relies strongly on experimentally measured quantities, and  
some new relevant results became available since the analysis presented in \cite{bas}. 
Furthermore, we  reanalysed the old data  
on charged multiplicities at low energies, which  in addition to some small errors propagated  
in the literature until now affected the result presented in \cite{bas}. 
 
As we already mentioned, the difference between the MLLA result and the experimental data on  
$\delta_{b\ell}$ would allow to probe the size of the next-to-MLLA effects of order  
$\alpha_s(M_b)N_{q\bar q}(M_b)$. 
 
Let us first take a fresh look at the MLLA expression for the charged multiplicity difference $\delta_{b\ell}$ 
\begin{equation} 
\delta_{b\ell}^{MLLA}=n_b^{dc} - N_{q\bar q}^{ch}(\sqrt{e} M_b) \label{eq7} 
\end{equation} 
in order to establish whether and where the two terms in the r.h.s. of Eq.(\ref{eq7}) 
require revision in the light of the current improvements in the understanding of experimental data. 
 
\medskip 
(i) Mean heavy hadron charged decay multiplicity, $n_b^{dc}$:\\ 
In the analysis of Ref.\cite{bas} the average number of charged particles coming from the decay of  
two $B$-hadrons was taken as  
$n_b^{dc}=2 N_b^{dc} = 11.0\pm 0.2$
In the present analysis we used the most recent result obtained from the  
combination of the ALEPH, CDF,  
DELPHI, L3, OPAL and SLD data on $B$-hadron production \cite{lepb},  
$N_b^{dc} = 4.955 \pm 0.062$,  
with an addition of  
$0.485 \pm 0.065$ tracks to include the charged decay products of $K^0_s$ and $\Lambda$,  
as measured by OPAL \cite{opalk}.  
There is still an issue of the role of heavier $B$-hadron states $(B^*, B^{**},\ldots)$ and on how fast  
the 'saturation' with rising energy can be established.  
Their contribution to the mean heavy hadron charged decay multiplicity,
$n_b^{dc}$ is usually evaluated  
with the help of Monte Carlo models.  
We used the value $0.22$ quoted by the SLD experiment~\cite{sld2}, which should be almost 
independent on $\sqrt{s}$ for c.m.s. energies above the $Z^0$ mass peak. 
 
We finally arrive at the value  
\begin{equation} 
n_b^{dc}= 11.10\pm 0.18 \label{eq9} 
\end{equation} 
which practically coincides with the previous result in \cite{bas}. 
 
\medskip 
(ii) Subtraction term $N_{q\bar{q}}^{ch}(\sqrt{e} M_b)$:\\ 
The second term in Eq. (\ref{eq7}) is related to the radiation within the 
dead cone, 
%
where primary gluons emitted off the $b$-quark and the $b$-quark itself 
act as a source of secondary soft radiation. 
In order to quantify the size of this term we have to address first the 
issue of the definition of the $b$-quark mass, which should be appropriate 
for the dead cone physics. 
 
As well known, within perturbative calculations it would be natural to take 
the pole in the quark propagator as the definition of the quark mass. 
By its very construction the pole mass is directly 
related to the concept of the free quark mass. However, due to the 
infrared effects the pole mass cannot be used with arbitrary 
high accuracy (see \cite{ckm} for recent review and references). 
Though in a more sophisticated calculation a mass definition which is less 
sensitive to the small momenta may appear to be more appropriate, the 
uncertainties in the quark mass of order of 
$\Lambda_{\rm QCD}$ are far beyond the accuracy of our consideration here. 
So for the purposes of this paper we use  the two-loop 
pole mass value, quoted in \cite{pdg}, 
\begin{equation} 
(M_b)_{pole}= 4.7 - 5.0 \  {\rm GeV} \label{eq10} 
\end{equation} 
which cover, in particular, some of the short distance mass prescriptions 
\cite{ahh,bigi1}. 
The scale $W_0^b= \sqrt{e}M_b$ at which the subtraction term $N_{q\bar{q}}^{ch}(W_0^b)$  
must be evaluated is then $\sqrt{s}= (8.0 \pm 0.25)$ GeV. 
 
Since there are no direct measurements of charged multiplicity  
at this energy, we estimate $N_{q\bar{q}}^{ch}$~(8 GeV)  
in the following way: 
 
\begin{itemize} 
 
\item use as many as possible experimental results on inclusive mean charged multiplicity  
$N^{ch}_{had}$ below and above $\sqrt{s}=8.0$ GeV, rather than restricting to a very  
limited energy range as in~\cite{bas}; 
 
\item fit the data points to evaluate $N^{ch}_{had}$(8 GeV) by interpolation, using different  
parameterisations and over a wide energy range,  
in order to test the consistency and 
stability of the results and to estimate a reasonably conservative uncertainty for  
$N^{ch}_{had}$(8 GeV); 
 
\item evaluate $N_{q\bar{q}}^{ch}$~(8 GeV) by subtracting the $c$-quark contamination  
from $N^{ch}_{had}$(8 GeV).  
 
\end{itemize} 
 
We studied all available data on mean charged particle multiplicity, $N^{ch}_{had}$, 
collected in $e^+e^-$ annihilations in the centre-of-mass energy range 1.4 GeV - 91 GeV.  
We considered only published results obtained in the 
continuum, thus away from the $J/\Psi$ and $\Upsilon$ resonances,  
which were determined following what is now considered a 
standard convention~\cite{lep1yr},  
namely including in the evaluation of the mean value all charged particles produced  
in the decays of particles with lifetimes shorter than $3\cdot 10^{-10}$ sec.  
\cite{gamma-gammam,mark1m,lenam,cleom,argusm,jadem,tassom,hrsm,tpcm,
amym,topazm,venusm,mark2m,alephm,delphim,l3m,opalm}. 
This means that the charged decay products of $K^0_s$ and of weakly decaying heavy-mesons  
(D,B$,\ldots$) and baryons ($\Lambda$,$\Sigma,\ldots$)  
as well as of their antiparticles  
must be considered, regardless of how far away from the interaction point the decay actually 
occurs. 
Unfortunately, some old publications, particularly those obtained at energies below 7 GeV,  
do not explain sufficiently well how the data were treated in this respect. 
We use only those ones which clearly considered at least charged decay products 
of $K^0_s$,  
that is known to be the dominant  contribution at low energies. 
Furthermore, we do not consider results obtained at energies which might suffer 
from  
threshold effects due to charmed meson pair production, including higher mass states,  
notably the data collected by MARK I~\cite{mark1m} in the interval 4.0 to 7.0 GeV. 
In order to evaluate $N^{ch}_{had}$(8.0 GeV) we fit the data points using the following  
parameterisations   
 
\begin{gather} 
{N^{ch}_{had} = a + b \cdot \ln (s) + c \cdot \ln^{2} (s)}, \label{log2} 
\\ 
{N^{ch}_{had} = a \cdot s^{b}}, \label{fermi} 
\\ 
{N^{ch}_{had} = a\cdot\alpha_{s}^{\beta}\cdot  
\exp(\gamma/\sqrt{\alpha_{s}}) }, \label{nlla} 
\end{gather} 
which are known to describe the data on mean charged  
multiplicity very well~\cite{opalm2}.
The parameters $a$, $b$ and $c$, as well as the effective scale $\Lambda$ 
not explicitly shown in (\ref{nlla}) but that enters the definition of the running 
coupling $\alpha_s$, are free parameters.\footnote{Note, that only (\ref{nlla}) 
is pQCD motivated, but we are using 
(\ref{log2}) and (\ref{fermi}) as well for interpolation purposes and error 
evaluation.} 
The two cases with three and five active flavours were considered in the 
calculation of $\alpha_s$ when making fits. 
 
We tested the consistency and the stability of the results by varying the fit energy 
range over the intervals: 
7 - 14 GeV; 7 - 44 GeV (to include the results from PEP and PETRA);  
7 - 62 GeV (to include results from TRISTAN) and 7 - 91.2 GeV  
(to include results from LEP1 and SLC), the common starting point of 7 GeV being well 
above the charmed meson production threshold. 
All mean multiplicities measured above $\sqrt{s}=10.5$~GeV were corrected 
for the effects caused by the b-quark.  
At each energy the correction was applied accounting for the fractions of the 
various quark species 
as predicted by the Standard Model and using the value $\delta_{b\ell}=3.1$ 
as measured experimentally. 
The total uncertainty associated with each data point was taken as the statistical 
and the systematic uncertainties added in quadrature.

 
All fits give a very good $\chi^2$, and the mean charged multiplicity predicted 
at $\sqrt{s}=8$~GeV is found to vary between 6.9 and 7.3. 
 
Our conclusion is that in the energy interval 
$\sqrt{s}\ = 7.75 - 8.25$~GeV 
%
%
\begin{equation} 
N^{ch}_{had}(\text{8.0 GeV}) = 7.1\pm 0.3\/. \label{eq11} 
\end{equation} 
The uncertainty includes the observed spread of values 
due to the choice of different parameterisations as well 
as the effect due to the uncertainty of the b-quark pole mass, $(M_b)_{pole}$. 
 
The result presented in (\ref{eq11}), however, refers to a mixture of 
$u,d,s,c$-events, while for the determination of $\delta^{MLLA}_{b\ell}$ 
from Eqs. (\ref{eq7}) and (\ref{eq9}) only the contribution to $N^{ch}_{had}$ 
from the light quarks $(q=u,d,s)$, $N^{ch}_{q\bar q}$, should be considered. 
In order to extract the light quark event multiplicity from (\ref{eq11}) 
we carefully studied the literature about the experimental results on the 
measurement of the multiplicity difference between the $q$- and $c$-quarks 
\begin{equation} 
\delta_{c\ell}^{exp} = N^{ch}_{c} - N^{ch}_{q\bar q}. 
\end{equation} 
 
 
At the time of the analysis of Ref. \cite{bas} only the results from  
MARK~II, TPC and TASSO were available. 
These results are affected by large uncertainties, and we also  
noticed in the literature some inconsistencies in the evaluation of  
$\delta_{c\ell}^{exp}$, which we corrected for. 
Much more precise results from OPAL~\cite{opal2} and SLD~\cite{sld2,sld3} are now available,  
and in the present analysis the experimental value of $\delta_{c\ell}^{exp}$ to be used  
for the correction was reevaluated, as discussed in detail in Appendix B.1. 
It is shown there, that the experimental results from 29 GeV to 91 GeV are well consistent  
with a constant value, and a weighted average yields 
\begin{equation}  
\delta_{c\ell}^{exp} = 1.0\pm0.4. \label{eq14} 
\end{equation} 
This value is about a factor two smaller than that used in \cite{bas},  
$\delta_{c\ell} = 2.2\pm1.2$, and is more precise.  
Since no direct measurements of $\dcl$ at $\sqrt{s} = 8$~GeV exist, we assume its 
constancy also at lower energies, as in \cite{bas}. 
Clearly, a direct and accurate measurement of $\dcl(\sqrt{s} = 8)$~GeV,  
for example by analysing radiative events with the proper effective energy at the 
BaBar and Belle experiments, would be much desirable to validate our hypothesis. 
 
We finally correct $N^{ch}_{had}$ for the effect of the 40\% admixture of $c\bar{c}$ events  
using this new result on $\delta_{c\ell}^{exp}$, and find for the  
light quarks 
\begin{equation} 
N_{q\bar q}^{ch} (\text{8.0 GeV})=6.7\pm0.34. \label{eq15} 
\end{equation} 
 
As a cross-check of this method, we estimate $N_{q\bar q}^{ch}$ also in the following way. 
Besides the b-quark contribution, we subtract also the c-quark contribution from all the  
mean charged multiplicities measured above the c-quark threshold. 
This is done using the value of $\delta_{c\ell}^{exp}$ presented in ({\ref{eq14}})  
and the Standard Model predictions for the c-quark fractions at each energy. 
 
With the exception of the MLLA parameterisation which in principle should not be used  
below the b-quark threshold,  
we can then extend the fitting procedure described above to the published results down to 1.4 GeV. 
This time the interpolation at 8 GeV provides directly the evaluation of $N_{q\bar q}^{ch}$, to be  
used for the calculation of $\delta^{MLLA}_{b\ell}$. 
The values of $N_{q\bar q}^{ch}$(8 GeV) are found to range in the interval 6.45 - 6.65,  
completely consistent with the value of $6.7 \pm 0.34$ quoted in ({\ref{eq15}}). 
 
We also compared our findings with the results from several global QCD fits to 
$N^{ch}_{had}$ at 8.0~GeV. 
The numerical solution of the MLLA evolution equation for the particle 
multiplicity generated by light quarks, supplemented by the full $O(\alpha_s)$ 
effects for $\epem$ annihilation \cite{lo}, gives $N^{ch}_{had}(\text{8.0 GeV}) = 6.5$. 
In this fit no effort was undertaken to separate the contributions from 
different flavours, so the fit which includes low energy data as well 
should be placed in between $N^{ch}_{had}$ and  $N^{ch}_{q\bar 
q}$, to be compared with (\ref{eq11}) and (\ref{eq15}). The 3NLO-fit 
\cite{dg} using the data above 10 GeV gives $N^{ch}_{had}(\text{8.0 GeV}) = 7.3$, 
consistent with (\ref{eq11}). 
Furthermore, a value $N^{ch}_{q\bar q}(\text{8.0 GeV} = 6.5$ is found by running the Pythia 6.2  
Monte Carlo program (in its default version) with light quarks only,  
with initial state radiation switched off  
and following the standard convention for the definition of  
mean charged multiplicity,%
\footnote{T. Sj\"ostrand, private communication.} 
in good agreement with our result (\ref{eq15}). 
 
Substituting (\ref{eq9}) and (\ref{eq15}) into Eq. (\ref{eq7}) 
we arrive at the revised MLLA expectation for the multiplicity difference 
\begin{equation} 
\delta_{b\ell}^{MLLA} = 4.4 \pm 0.4  \label{eq16} 
\end{equation} 
which is $\sim 1.0$ unit lower than the result reported in \cite{bas} 
and has half of its uncertainty. 
 
The comparison of the MLLA result (\ref{eq16}) with the available  
experimental data on  
$\delta_{b\ell}$ in $\epem$ annihilation is shown in Fig. 2; here we 
included also the  
reevaluated results of  
DELCO, MARK II, TPC, TASSO, TOPAZ and 
VENUS (see Appendix B.2). The new experimental average 
is given by 
\begin{equation}
\delta_{b\ell}^{exp}=3.14\pm0.14.
\label{delexp}
\end{equation} 
\begin{figure}[t!]
\begin{center} 
\mbox{\epsfig{file=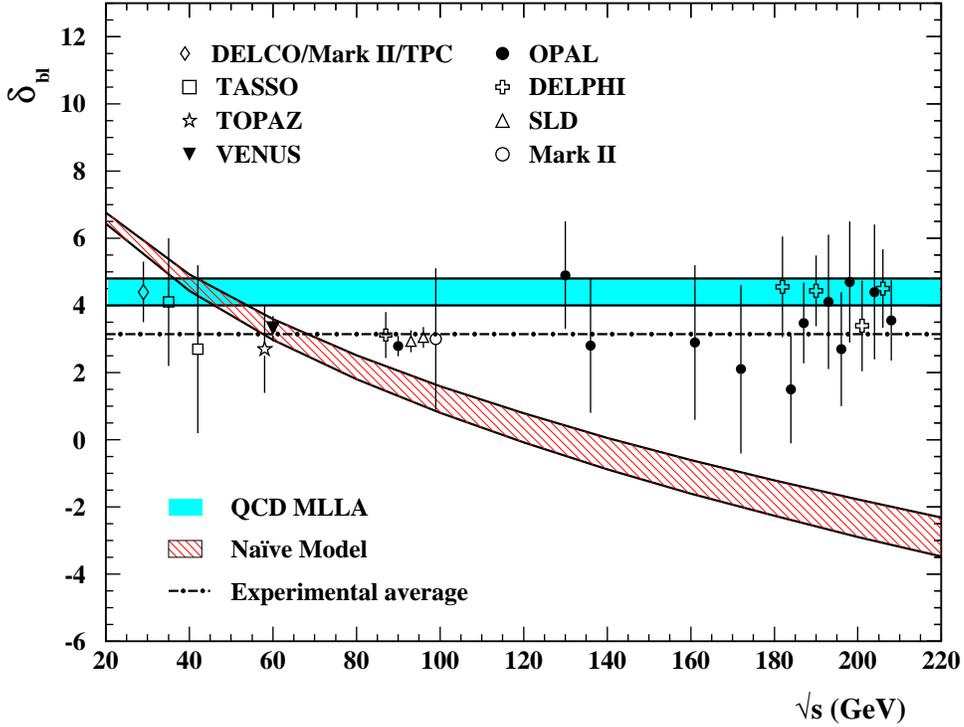,bbllx=1.0cm,bblly=7.cm,bburx=20.5cm,%
bbury=22.cm,width=13.0cm}} 
\end{center} 
\vspace{-0.3cm} 
\caption{ 
Experimental measurements of $\delta_{b\ell}$ plotted as a function 
of the c.m.s. energy, $\protect\sqrt{s}$; data below 90 GeV reevaluated 
(see Appendix B.2).  
The revised MLLA expectation using $\delta_{b\ell}^{MLLA}=4.4\pm0.4$  
is indicated by the shaded 
area. Also shown is the ``na\"\i ve model'' \protect\cite{pcr}
 based on the reduction of energy scale 
(dashed area).}  
\label{fig2} 
\end{figure} 
We can say that, qualitatively,  the previous conclusion  
that the experimental mean value is lower 
than the absolute value of the MLLA  
prediction remains valid. 
Quantitatively, however, the agreement between the data and the theory  
definitely improves. 
 
Finally, we turn to the question of whether the remaining discrepancy can be 
attributed to the next-to-MLLA contributions. 
First we note that the experimental value of the multiplicity difference
$N_{q\bar q}(W)-N_{b\bar b}(W) = n_b^{dc}-\delta_{b\ell}^{exp}=7.96\pm 0.23$ 
and the MLLA
expectation $N_{q\bar q}(\sqrt{e} M)$ as evaluated in (\ref{eq15}) 
differ by a relative amount $< 20\%$ which is 
of the order of the expected correction term in (\ref{eq3}) of
$\cO{\alpha_s(M)}$.
To gain insight at the quantitative level, we consider
first the size of the above multiplicity difference in the Double
Logarithmic Approximation (DLA). This is given by  (\ref{pi2final})
but with the r.h.s. replaced simply by $N_{q\bar q}(M)$ if the dominant
contribution to $N_0$ in Eq. (\ref{memu2}) is taken. For
 $b$-quark jets, this
requires evaluation of the multiplicity at $M_b\sim 4.85$ GeV. We estimate 
	$N^{ch}_{had}(M_b)\approx 5.1$ ($N_{q\bar q}^{ch}(M_b)\approx 4.7)$ 
and, therefore, $\delta_{b\ell}^{DLA}\sim 6.4$. 
This is about two units above the MLLA prediction (\ref{eq16}) which, in turn, is 
about one unit above the data in (\ref{delexp}) indicating convergence.  
 
In the next-to-MLLA two large ``$\pi^2$-contributions'' are derived 
explicitly, see Eq. (\ref{pi2final}) in Appendix A. The final  
expression involves 
the coupling at scale $M$ which we derive from the 1-loop formula 
with $\Lambda=250$ MeV, as typically used in MLLA applications (see 
for example, Ref \cite{ko,dkmt2}),  and we obtain $\alpha_s(M_b)=0.23$ for 
$n_f=3$ flavours. Then from (\ref{pi2final}) we find  
$N^{ch}_{q\bar q}(W)- N^{ch}_{Q\bar Q}(W)= 
N^{ch}_{q\bar q}(\sqrt{e}M)\times 1.27\approx 8.5$. This finally gives 
the result including 
these next-to-MLLA contributions $\delta_{b\ell}\approx 2.6\pm0.4$. 
We, therefore, conclude, that the MLLA prediction is already close to the 
experimental data in (\ref{delexp}), 
and the remaining difference is of the order of the expected next-to-MLLA 
contributions.

 
 
\subsection{Results on charm quark jets} 
Since the scale relevant for the charm quark,  
$W_0^c\sim\sqrt{e}M_c$, is significantly lower than in the $b$-quark  
case the predictions are less reliable. 
 
The two-loop $c$-quark pole mass is quoted in \cite{pdg} as 
\begin{equation} 
(M_c)_{pole}= 1.47-1.83\ \GeV. 
\end{equation} 
We evaluated the size of the subtraction term $N^{ch}(W_0^c)$ where we 
followed 
the same strategy as described above, restricting the  
multiplicity fits to the energy range  1.4  - 10.45 GeV. 
The predicted value at $W_0^c=2.7$ GeV is found to vary between 3.5 and 3.9, and 
we arrive at $N_{\qq}^{ch}(2.7\ \GeV) = 3.7\pm 0.3$. 
Using the $c$-quark decay multiplicity 
$n_c^{dc}= 5.2 \pm 0.3$ we obtain the MLLA expectation for the charged particle multiplicity  
difference in the $c$-quark case  
\begin{equation} 
\delta_{c\ell}^{MLLA} = 1.5 \pm 0.4 
\label{deltacl} 
\end{equation} 
which is basically the same as the previous number $\delta_{c\ell}^{MLLA} = 
1.7 \pm 0.5$ in \cite{bas}. 
The result (\ref{deltacl}) is consistent with the new more precise 
experimental average given by Eq.~(\ref{eq14}). As in the 
case of $\delta_{b\ell}$ the 
theoretical MLLA result lies now above the experimental value which is  
expected due to the presence of the higher order effects.  
 
We also note an 
interesting aspect of the difference between the $b$ and $c$ quark multiplicities 
$\delta_{bc}=\delta_{b\ell}-\delta_{c\ell}$. 
Since the $M$-dependence of the next-to-MLLA term in (\ref{pi2final})  
is weaker than that of the 
leading  $N_{q\bar q}(\sqrt{e}M)$ contribution, the 
multiplicity difference $\delta_{cb}=\delta_{b\ell}-\delta_{c\ell}$ is less 
affected by this correction, and can be better approximated by the MLLA 
result.  
If we compare the experimental  
and theoretical numbers obtained from the results derived above  
\begin{equation} 
\delta_{bc}^{MLLA}=2.9\pm0.6,\qquad  
\delta_{bc}^{exp}=2.1\pm0.4, 
\end{equation} 
we find indeed that, contrary to the difference $\delta_{b\ell}$,  
within the slightly larger errors, there is a reasonable agreement 
between the data and the MLLA prediction for this multiplicity difference 
involving $b$ quarks. 
 
 
\section{Conclusions} 
The comparison of particle multiplicities in heavy and light-quark initiated 
jets provides a specific test of the perturbative approach to multiparticle 
production. In this approach the particle multiplicities  
in $\epem$ annihilation are directly 
proportional to the gluon multiplicities generated by multiple successive  
bremsstrahlung processes from the primary quarks.  
In the case of a primary heavy quark the 
small angle radiation is kinematically suppressed (dead cone 
effect). Also the subsequent gluon emission is affected by the mass 
effects, 
which results in the loss of coherence of soft gluon radiation 
off the heavy quark and the primary gluon. 
The result can be represented as an appropriate expansion  
in $\sqrt{\alpha_s}$  where the leading double logarithmic and 
next-to-leading (MLLA) terms have been known for quite a while, whereas certain large 
contributions in the next-to-MLLA order are discussed here.  
 
The main aim of this study is to sharpen the tests of the perturbative approach 
by accounting for all currently available data on $\epem$ annihilation. 
More accurate theoretical predictions for the difference of multiplicities in 
light and heavy quark jets are obtained.  
The expected energy independence of this difference is nicely confirmed. 
The same difference in 3-jet events is expected to agree with that in 2-jet events 
at the corresponding $q\bar q$ c.m.s. energy, and this is supported by preliminary data. 
As compared to the previous analysis, the updated MLLA prediction for the  
absolute value of the multiplicity  
difference comes closer to the experimental data. It is shown 
that the remaining difference is of the order of the next-to-MLLA 
corrections considered.  
This way the specific effects related to soft gluon bremsstrahlung 
from heavy quarks and their impact on the generation of the gluon cascade  
are reaching quantitative understanding within the perturbative approach. 
 
It would be very interesting to extend the measurements of $\dbl$ and 
$\dcl$ to lower energies, for example down to the region
accessible at the B-factories. 
In particular, a direct measurement of $\dcl$ at $\sqrt{s}= 8.0$~GeV,
for example from the analysis of events with initial state photon radiation,
 would be very important to confirm our assumption that
$\dcl$ remains constant below 29 GeV. 
Further tests of the QCD predictions at higher 
energies at a future linear collider will be interesting. 
 
\section*{Acknowledgement} 
 
We thank Pedro Abreu, Alessandro De Angelis, Bill Gary, Klaus Hamacher, 
 Andre Hoang, Martin Siebel, Torbj\"orn  Sj\"ostrand, 
Sergey Troyan and Nikolai Uraltsev for useful discussions. 
VAK thanks the Theory Group of the Max-Planck-Institute, Munich for the hospitality. 
 
\newpage 
 
\def\theequation{\Alph{section}.\arabic{equation}} 
\begin{appendix} 
\setcounter{equation}{0} 
\setcounter{section}{1} 
\section*{Appendix A} 
\label{app:A} 

\subsection{Single gluon emission in MLLA and beyond} 
The exact first order expression for probability of single gluon 
emission off the heavy quark pair can be written in the following 
form \cite{dktHQ,dks}, in analogy with QED \cite{bk}, 
\beeq 
\label{wgeneralzeta} 
{dw_V} = \frac{C_F\as}{\pi\,v}\> \frac{dz}{z}\> 
\frac{d\eta}{\sqrt{1-\eta}} \left\{\, {2(1-z)} \> 
  \frac{\eta-\eta_0}{\eta^2} \>\>+\>\> z^2\left[\,\frac{1}{\eta}- 
    \frac12 \,\right] \zeta_V^{-1} \,\right\} \,, \eeeq 
with $z$ the gluon energy fraction and $\eta$  
an angular variable and 
\beeq  
 1 \>\ge \> \eta =1-\beta^2\cos^2\Theta_c \>\ge\>\eta_0= 
 \frac{4m^2}{1-z}\,,\quad m\equiv \frac{M}{W}\ll 1\,,  
\eeeq 
where $\beta$ is the quark velocity and 
$\Theta_c$ is the polar gluon angle in the \QQ c.m.s.,  
\beeq 
\beta^2=\beta^2(z)= 1-\frac{4m^2}{1-z} \> \le\> v^2=1-4m^2 \> \ge\> 
z\,. 
\eeeq  
The first term in curly brackets in \eqref{wgeneralzeta} contains the 
main (double logarithmic) contribution and corresponds to universal 
soft gluon \br. In accordance with the Low-Barnett-Kroll theorem 
\cite{LBK}, both $dz/z$ and $dz$ parts of the radiation 
density have a classical origin and are, therefore, universal, independent 
of the process (and of the quark spin).  This term explicitly exhibits the 
dead cone phenomenon: ``soft'' radiation vanishes in the forward 
direction, $\sin\Theta_c\to 0$, $\eta\to\eta_0$. 
 
The second term proportional to $dz\,z$ (hard gluons) depends, 
generally speaking, on \QQ production mechanism.  Namely, both $-1/2$ 
subtraction term and the factor $\zeta_V= (3-v^2)/2 = 1+2m^2$ would be 
different for production current other than the vector current.  We 
include this remark to stress that at the level of ${\as}$ corrections 
(as well as of power suppressed effects $\cO{\sqrt{\as}m^2}$) the mean 
multiplicity acquires process-dependent contributions from 3-jet 
ensembles and cannot be treated any longer as an intrinsic 
characteristic of the \QQ system. 
 
To obtain the mean parton multiplicity with the MLLA accuracy it 
suffices to supply (\ref{wgeneralzeta}) with the gluon cascading 
factor which depends, together with the running coupling, on the 
argument  
\begin{equation}  
  k_t^2 = \left(\frac{zW}{2}\right)^2 \, \eta\,,  
\end{equation} 
see \cite{dktHQ}.  Neglecting relative corrections $\cO{\as}$ and 
$\cO{m^2}$ we obtain for the mean multiplicity 
\begin{equation}\label{memu1} 
 N_{\QQ}(W)= N_0 - N_1\,, 
\end{equation} 
where 
 
\begin{eqnarray} 
\label{N0} 
N_0 &=& \frac{C_F}{\pi}\int_0^{v^2} \frac{dz}z \int_{\eta_0}^1 
\frac{d\eta}{\eta}\,\frac{1+(1-z)^2}{\sqrt{1-\eta}} 
\cdot \left[ \as \cN_G\right]\left(k_t\right) , \\ 
\label{N1} 
N_1 &=&   \frac{C_F}{\pi}\int_0^{v^2} \frac{dz}z \int_{\eta_0}^1 
\frac{\eta_0\, d\eta}{\eta^2}\>\> 
2(1-z)  
\cdot \left[ \as \cN_G\right]\left(k_t\right). 
\end{eqnarray} 
We start by analysing the leading term \eqref{N0}. 
 
\paragraph{${\bf N_0}$.} 
The kinematical factor $1/\sqrt{1-\eta}$ somewhat enhances the 
contribution of the large angle region, $\eta=\cO{1}$, and should be 
taken into consideration in the leading DL term.  The corresponding 
(SL) correction can be approximately accounted for by pushing up the 
upper limit of the logarithmic integration.  Indeed, given that the 
factor $\cF\equiv\as\cN$ depends on $\eta$ logarithmically, the chain 
of approximations follows: 
\beeq \int_{\eta_0}^1 \frac{d\eta}{\eta\sqrt{1-\eta}}\cF(\eta) = 
\int_{\eta_0}^1 \frac{d\eta}{\eta}\cF(\eta) + \int_{\eta_0}^1 
\frac{d\eta}{\eta}\left(\frac1{\sqrt{1-\eta}}-1\right) 
\cF(\eta) \nonumber \\ 
\approx \int_{\eta_0}^1 \frac{d\eta}{\eta}\cF(\eta) \>+\> \ln 4\cdot 
\cF(1) \approx \int_{\eta_0}^4 \frac{d\eta}{\eta}\cF(\eta)\,.   
\eeeq 
It is straightforward to check that  
the omitted terms are limited from above by  
the $\cO{\as(W)}$ and $\cO{m^2}$ terms. 
Natural rescaling of the integration variable, $t=\eta W^2/4$, leads 
to 
\begin{equation} \label{N0sub} 
 N_0 = \frac{C_F}{\pi}\int_0^{v^2}dz \frac{1+(1-z)^2}{z} 
 \int_{M^2/(1-z)}^{W^2} \frac{dt}{t}\, \left[ \as 
 \cN_G\right]\left(k_t\right), 
\end{equation} 
with 
\begin{equation}\label{ktdef} 
 k_t^2 = z^2\, t\>. 
\end{equation} 
Now we represent \eqref{N0sub} as  
\begin{equation} \label{memu2} 
 N_0 \>=\> N_{\qq}(W) - N_{\qq}(M) -N_2\,, 
\end{equation} 
where we have singled out an (enhanced) next-to-MLLA correction term 
that we will consider later, 
\begin{equation}\label{N2}  
 N_2 =N_2(M)\>=\>  \frac{C_F}{\pi}\int_0^{v^2}dz 
 \frac{1+(1-z)^2}{z} \int^{M^2/(1-z)}_{M^2} \frac{dt}{t}\, 
 \left[\as \cN_G\right]\left(k_t\right), 
\end{equation}  
and introduced the function  
\begin{equation}\label{Nlight} 
 N_{\qq}(W) \>=\> \frac{C_F}{\pi}\int_0^{1}dz \frac{1+(1-z)^2}{z} \int^{W^2} 
\frac{dt}{t}\, \left[ \as \cN_G\right]\left(k_t\right)  
\end{equation} 
that describes the {\em light quark}\/ event multiplicity at the 
c.m.s.\ energy $W$.  
 
We observe that the dead cone suppression naturally results in the 
expression for the accompanying multiplicity in \QQ events as a 
difference of light quark multiplicities at c.m.s.\ energies $W$ and 
$M$.   
The MLLA correction $N_1$ defined in \eqref{N1} modifies the 
effective energy of the subtraction term $N(M)$ in \eqref{memu2}.

\paragraph{${\bf N_1}$.} 
 
Since the $\eta$ integral in \eqref{N1} is non-logarithmic and is 
concentrated in the region $\eta\sim\eta_0\ll 1$, we allowed ourselves 
to drop the $1/\sqrt{1-\eta}$ factor here as producing a negligible 
$\cO{m^2}$ correction.  
We have  
\begin{equation}\label{dN1}  
 N_1 = \frac{C_F}{\pi} \int_0^1 dz\,\frac{2(1-z)}{z} 
 \int_1^\infty\frac{du}{u^2} \left[ \as \cN_G\right](k_{t0}) \,, \quad 
 k_{t0}^2 = \frac{z^2}{1-z}M^2\, u\,.  
\end{equation}  
Though the {\em collinear}\/ logarithmic enhancement disappears here, the 
{\em soft}\/ one is still present (contrary to $N_2$) and promotes 
$N_1$ to the $\sqrt{\as}$ (MLLA) level. 
 
First we observe that the $(1-z)$ rescaling of the argument of the 
cascading factor $[\as\cN]$ is negligible as it produces a
$\cO{\as^{3/2}}$ correction (next-to-next-to-MLLA).  Then, making use of the expansion 
$$ 
 \int_1^\infty \frac{du}{u^2} \, F(\ln u) \>=\> F(0) +F'(0)+\ldots 
$$ 
 and replacing the factor $2(1-z)$ by the numerator of the full 
 quark $\to$ gluon splitting function, $1+(1-z)^2$, we arrive at 
\begin{equation} \label{N12} 
 N_1 = \frac{C_F}{\pi}\left( \int_0^{1}dz \,\frac{1+(1-z)^2}{z} 
  [\as\cN_G](z\,\sqrt{e}M) -\frac12 [\as\cN_G](M)\right), 
\end{equation}  
which holds with the next-to-MLLA accuracy (including 
$\cO{\as}$).  By comparing \eqref{N12} with \eqref{Nlight} and 
recalling \eqref{ktdef} we can express the MLLA correction as 
the {\em logarithmic derivative}\/ of the light quark multiplicity: 
\begin{equation}  
 N_1 \>=\> \frac12\, N'_{\qq}(\sqrt{e}M)\,; \qquad N'_{\qq}(Q)\equiv 
 \frac{d}{d\ln Q}\, N_{\qq}(Q)\,.   
\end{equation} 
Invoking \eqref{memu1} and \eqref{memu2}, for the \QQ event 
multiplicity we finally obtain 
\beeq \label{NwithN2} 
 N_{\QQ}(W) &=& N_{\qq}(W) - \left[\, N_{\qq}(M) + 
   \frac12\,N'_{\qq}(\sqrt{e}M)  
+ \ldots\, \right] - N_2 \nonumber\\ 
&\simeq&  N_{\qq}(W)-N_{\qq}(\sqrt{e}M) \,-\,N_2\,. 
\eeeq 
This proves the MLLA subtraction formula \eqref{eq3}.

\paragraph{${\bf N_2}$.} 
 
The effect due to the $(1-z)$ rescaling of the lower limit of the 
$t$-integration in \eqref{N0sub} produces a $\pi^2$ enhanced next-to-MLLA 
correction $N_2$. We have 
\beeq  
 N_2(M) &=& \frac{C_F}{\pi}\int_0^{v^2}dz 
 \frac{1+(1-z)^2}{z} \int^{M^2/(1-z)}_{M^2} \frac{dt}{t}\, 
 \left[\as \cN_G\right]\left(k_t\right) \nonumber \\ 
& \simeq& \frac{C_F}{\pi} \int_0^1 dz \frac{1+(1-z)^2}{z} 
 \ln\frac1{1\!-\!z} \left[ \as \cN_G\right]\left(M\right) \nonumber\\ 
& =& 
 \frac{C_F}{\pi} \left(\frac{\pi^2}{3}-\frac54\right) 
 \left[\as\cN_G\right](M) \cdot \left\{1+\cO{\as^{1/2}(M)}\right\}.   
\eeeq  
In terms of the event multiplicity ($N_{\qq}\simeq 2C_F/N_c\cdot 
\cN_G$) we arrive at the relative correction 
\begin{equation} \label{N2rel} 
 \frac{N_2(M)}{N_{\qq}(M)} \>\simeq\> \frac{N_c\as}{2\pi} \cdot 
 \left(\frac{\pi^2}{3}-\frac54\right).   
\end{equation}

\subsection{Two gluon (dipole) correction} 
 
To derive the probabilistic MLLA equations describing parton cascades one 
has to analyse, in particular, ensembles of many energy ordered gluons 
radiated at arbitrary angles and demonstrate that, after having 
taken into full account multiple interference diagrams, one arrived at 
the pattern of {\em angular ordered}\/ (AO) successive gluon 
emission~\cite{dkmt2}. 
Reduction of interference graphs to the probabilistic AO scheme is not 
exact: there is a ``remainder''. In particular, the first such remainder 
appears at the $\as^2$ order and describes radiation of two soft 
gluons (with energies $k_2\ll k_1\ll W$) at large angles with respect 
to the \qq pair and to each other.  
The angular structure of the remainder $R^{(2)}$ is as follows  
\begin{equation}\label{R2}  
 R^{(2)} \>=\> C_FH_+^1\cdot N_cD^2_{-[+1]}+ C_FH_-^1\cdot N_cD^2_{+[-1]} \,, 
\end{equation} 
where $\pm$ mark the momenta of $q$ and $\bar{q}$,  
the factor $H$ describes independent gluon emission, 
\begin{equation}  
H^i_{\ell} \> =\> \frac2{a_{i\ell}}\,, \qquad 
a_{ik}=q^2\frac{(p_i\cdot p_k)}{(p_i\cdot q)(p_k\cdot q)}= 1-{\bf 
  n}_i{\bf n}_k= 1-\cos\Theta_{ik}\,,  
\end{equation} 
and $D$ is the so-called ``dipole factor'', 
\beeq\label{CPS22}  
 D^i_{\ell\,[mn]} &\equiv& I^i_{\ell m}-I^i_{\ell n}\,, \\ 
 I^i_{\ell m} &=& \frac{a_{i\ell}+a_{im}-a_{\ell m}}{a_{i\ell}a_{im}} \,. 
\eeeq 
The dipole remainder possesses no collinear singularities,  
\begin{equation}\label{CPS28} 
 \int\frac{d\Omega_2}{4\pi}\int\frac{d\Omega_1}{4\pi}\> H_+^1\> 
 D^2_{-[+1]} = \int_0^1 \frac{dx}{1-x}\ln x = - \zeta(2) = 
 -\frac{\pi^2}{6}\,,   
\end{equation} 
so that the integration of \eqref{R2} over the gluon angles gives 
\begin{equation}  
 \int\frac{d\Omega_2}{4\pi}\int\frac{d\Omega_1}{4\pi}\> R^{(2)}({\bf n}_2,{\bf n}_1) 
 = 2C_FN_c\cdot  \left(-\frac{\pi^2}{6}\right). 
\end{equation} 
With account of the gluon cascading factor, logarithmic integrals over 
the gluon energies induce the next-to-MLLA correction to the event 
multiplicity 
\begin{equation}\label{CPS58}  
 \Delta N_{\qq}(Q) = 2C_F N_c\left(-\frac{\pi^2}{6}\right)\int^Q 
 \frac{dk_1}{k_1}\>\frac{\as}{2\pi}\int^{k_1}\frac{dk_2}{k_2} 
 \>\frac{\as}{2\pi}\>\cN_G(k_2)\,. 
\end{equation} 
Now we estimate the energy integrals using 
$$ 
\int^{k}\frac{dk'}{k'} \>\cN_G(k') \>\simeq\> \frac1{\gamma_0}\cdot 
\cN_G(k)\,, \qquad \gamma_0= \sqrt{\frac{2N_c\as}{\pi}}\,, 
$$ 
with $\gamma_0$ the DLA multiplicity anomalous dimension,   
and obtain another $\pi^2$ enhanced relative correction  
\begin{equation}\label{dipest}  
\frac{\Delta N_{\qq}(Q)}{N_{\qq}(Q)}\>=\>  
 -N_c^2\,\frac{\pi^2}{6}\frac{(\as/2\pi)^2}{\gamma_0^2} \>=\>  
 -\frac{N_c\as(Q)}{2\pi} \cdot \frac{\pi^2}{24} \,.   
\end{equation} 
This means that the true multiplicity 
$N_{\qq}=N_{\qq}^{(\mbox{\scriptsize MLLA})} + \Delta N_{\qq}$ and its 
MLLA estimate are related as follows 
\begin{equation} \label{related} 
N_{\qq}^{(MLLA)}(Q) \>\simeq\> N_{\qq}(Q)\cdot \left(1 + \frac{N_c\as(Q)}{2\pi} 
  \cdot \frac{\pi^2}{24}\right). 
\end{equation} 
Now we return to the expression \eqref{NwithN2}.  
\begin{equation} 
 N_{q\bar q}(W) \,-\, N_{\QQ}(W) \>=\>  N_{q\bar q} (\sqrt{e}M)  
\>+\>N_2(M)\,. 
\label{eq32} 
\end{equation} 
The first observations we make is that in the difference $ N_{\qq}(W) 
\>\,-\, N_{\QQ}(W)$ the two-gluon dipole corrections cancel since, as 
we discussed above, large angle soft gluon emission is 
insensitive to quark mass. Therefore, we can look upon the l.h.s.\  
as being constructed of the true multiplicities. 
 
On the contrary, the factor $N_{\qq}$ on the r.h.s.\ of \eqref{eq32} 
is the theoretical (MLLA) expression. Relating it with the true 
multiplicity via \eqref{related} results in 
\begin{equation} \label{pi2final} 
  N_{q\bar q}(W) -  N_{\QQ}(W) \,=\,  N_{q\bar q} (\sqrt{e}M) 
  \cdot \left\{ 1+ \frac{N_c\as(M)}{2\pi} 
\left[\, \frac{\pi^2}{24} + \left(\frac{\pi^2}{3}-\frac54 \right)\right] \right\}, 
\end{equation} 
where we inserted the expression \eqref{N2rel} for the first enhanced 
correction $N_2$. Numerically, the first term in the square bracket 
from the dipole corrections at large emission angles  
amounts only to about 4\% whereas the second one, 
which improves the description of  
the small angle emission from the heavy quark, is about 5 
times larger. The result (\ref{pi2final}) is not  
claimed to be complete at this order but it   
includes the important $\pi^2$ contributions considered to be dominant 
and  shows the size of the next-to-MLLA terms. 
Remarkably, both corrections work in the same direction increasing the  
difference between the light and heavy quark companion multiplicities.  
 
 
\setcounter{equation}{0} 
\setcounter{section}{2}  
\setcounter{subsection}{0} 
 
\section*{Appendix B} 
\label{app:B} 
 
\subsection{On the measurement of $\delta_{c\ell}$} 
\label{dcl} 
 
The experimental determination of $\delta_{c\ell}$ at different energies  
is very important for this analysis. 
As discussed in Sect. 3.2, a key point in the evaluation of the absolute value of 
the MLLA prediction for $\delta_{b\ell}$ is the determination of the light-quark  
mean multiplicity, ${N}^{ch}_{q\bar{q}}$, at $\sqrt{s}=$~8 GeV. 
Experimentally one measures the mean charged multiplicity of an unbiased inclusive sample 
of hadronic events, ${N}^{ch}_{had}$, and then subtracts the contamination of  
heavy-quark-initiated events. 
This can be done if one knows the fractions of light and heavy-quark events present  
in the sample, $f_{\ell}$ and $f_Q$, as well as the difference between mean 
multiplicities  
of the heavy and light quarks, $\delta_{Q\ell}$, using the relation 
\begin{equation}  
 {N}^{ch}_{had}\, =\, f_{\ell} \cdot {N}^{ch}_{q\bar{q}} +  
f_c \cdot ({N}^{ch}_{q\bar{q}} +  
 \delta_{c\ell}) +  f_b \cdot ({N}^{ch}_{q\bar{q}} +\delta_{b\ell}).   
\end{equation} 
 
At $\sqrt{s}=$~8 GeV, where only the $c$-quark-initiated events are produced  
on top of the 
light-quark events, a direct measurement of $\delta_{c\ell}$ is not available 
and, thus, its value must be evaluated from the knowledge of experimentally measured  
values of $\delta_{c\ell}$ at different energies.     
Moreover, the knowledge of  $\delta_{c\ell}$ is necessary to derive $\delta_{b\ell}$  
from the results of those experiments which do not measure directly the c-quark  
event mean multiplicity. 
The measurement of $\delta_{c\ell}$ is difficult because it is not easy to  
select experimentally a highly enriched sample of c-quark initiated events. 
 
So far, only five experiments published their results on the direct measurement of  
the mean charged particle multiplicities, ${N}^{ch}_{c}$ and 
${N}^{ch}_{\ell}$, for $e^+e^- \rightarrow c\bar{c}$  
and $e^+e^- \rightarrow \ell\bar{\ell}$ ($\rm \ell\bar{\ell} \equiv q\bar q  = \light$)  
events, including the evaluation of statistical and systematic uncertainties:  
MARKII~\cite{row} and TPC~\cite{tpc} at $\sqrt{s}=$ 29 GeV, TASSO~\cite{tasso1,tasso2}  
at $\sqrt{s}=$ 35 GeV and OPAL~\cite{opal2} and SLD~\cite{sld2,sld3} at $\sqrt{s}=$ 91.2 GeV. 
These results, together with the derived values of $\delta_{c\ell}$  
and their weighted average are presented in Table 1.\footnote{There are published results
on the measurements of $\nc$ and $\nl$ also at LEP2 
energies~\cite{delphi2,opal3,delprel}.
Unfortunately, the limited statistics available at each energy did not allow 
the efficient $c$-quark tagging, comparable to that at the $Z^0$ peak.
Therefore, the selection of highly enriched c-quark samples was not possible.
As a consequence the measurements of $\nc$ are affected by large uncertainties,
and cannot be used for a meaningful evaluation of $\dcl$.}  
 
It should be mentioned that the two results from SLD~\cite{sld2,sld3} were obtained from  
two completely independent event samples.  
The most recent one was collected with an upgraded detector, using a different  
experimental procedure and with different sources of systematic errors. 
We then consider the two results practically uncorrelated.   
It should also be noticed that the results of MARKII and TPC presented in Table 1 
are different from those derived in~\cite{bas}, and used to evaluate the 
light-quark charged mean multiplicity at $\sqrt{s}=$~8 GeV in the same article. 
This is simply due to the fact that in~\cite{bas} the values of $\nl$ used 
to calculate $\dcl$ were not those quoted in the publications~\cite{row} and~\cite{tpc}, 
but they were recalculated assuming a common mean value for the  
total average multiplicity, $N^{ch}_{had}$, as determined by different experiments
at energies surrounding $\sqrt{s} \approx$~29~GeV. 
This procedure was meant to reduce the uncertainty on the derived values of $\dcl$ 
and $\dbl$ but is rather dangerous since information about 
the strong correlations among 
$N^{ch}_{had}$, $\nl$, $\nc$ and $\nb$ existing within the same measurement,  
is lost if one considers a mean value over different experiments for only one of these variables.  
That is why we rather preferred to use the published results which were all obtained within  
the same measurement.   

\begin{table}[t] 
\begin{center} 
\begin{tabular}{|c||c|c|c|c|} 
\hline 
 Experiment & $\sqrt{s}\ (GeV)$  & $\nc$ & $\nl$   & $\dcl$             \\ 
\hline 
\hline 
 
MARKII~\cite{row} & 29 & 13.2 $\pm$ 1.0 & 12.2 $\pm$ 1.4 & 1.0 $\pm$ 1.7 \\ \cline{2-5} 
TPC~\cite{tpc}    & 29 & 13.5 $\pm$ 0.9 & 12.0 $\pm$ 0.9 & 1.5 $\pm$ 1.3 \\ \cline{2-5} 
TASSO~\cite{tasso1,tasso2} & 35 & 15.0 $\pm$ 1.2 & 11.9 $\pm$ 1.2 & 3.1 $\pm$ 1.6 \\ \cline{2-5} 
OPAL~\cite{opal2}   & 91.2 & 21.52 $\pm$ 0.62 & 20.82 $\pm$ 0.44 & 0.69 $\pm$ 0.62 \\ \cline{2-5} 
SLD~\cite{sld2}      & 91.2 & 21.28 $\pm$ 0.61 & 20.21 $\pm$ 0.24 & 1.07 $\pm$ 0.59 \\ \cline{2-5} 
SLD~\cite{sld3}     & 91.2 & 21.096 $\pm$ 0.653 & 20.048 $\pm$ 0.316 & 1.048 $\pm$ 0.718 \\ \cline{2-5} 
\hline 
\hline 
Average             &      &                    &                    &  1.03 $\pm $ 0.34 \\ \cline{2-5} 
\hline 
\end{tabular} 
\end{center} 
\caption{\label{table1} 
Mean charged particle multiplicities,  $\nc$ and $\nl$, for $\cc$ and $\rm \ell\bar{\ell}$ 
($\rm \ell\bar{\ell} = \light$) events and the difference $\dcl = \nc - \nl$, measured 
at different energies. The results are corrected for detector effects as well as for initial
state radiation effects.  
Charged decay products from $K^o_S$ and $\Lambda$ decays are included.  
We derived $\nl$ for TASSO from the published values of $\nb$, $\nc$ and 
${N}^{ch}_{had}$, assuming the Standard Model quark fractions. 
The quoted errors are obtained by  
combining the statistical and the systematic errors in quadrature. 
OPAL and SLD errors on $\dcl$ were published considering also correlations.  
The weighted average assumes no correlations among the various experimental results.} 
\end{table} 
%
%
There are two more experimental results on $\nc$ and $\nl$ published in the literature,  
one by the HRS collaboration at $\sqrt{s}=29$~GeV~\cite{hrs} and one by the DELPHI  
collaboration at $\sqrt{s}=91$~GeV~\cite{delphi1}. 
Unfortunately, only the statistical uncertainties were evaluated in these analyses, and
since the contribution of the systematic errors to the total error quoted by the other 
experiments is important, or even dominant, we did not consider the results from HRS 
and DELPHI in our weighted average. 
We show in the following that in any case, under reasonable assumptions about 
the size of the total errors, the final result would not change significantly if we did.   
The DELPHI experiment measured $\nc$, $\nb$ and $\nl$ and found $\dcl = 1.64$.  
The total uncertainty (statistics and systematics combined) on $\dbl$ quoted  
in their analysis is about a factor two larger than those quoted by  
SLD~\cite{sld2,sld3} and OPAL~\cite{opal2}, and if we assume a similar relative 
precision also for $\dcl$, by comparison with the SLD and  
OPAL total uncertainties we get for DELPHI $\dcl = 1.64 \pm 1.2$. 
Our weighted average in Table 1 would change to $\langle\dcl\rangle = 1.07 \pm 0.33$ if we would  
include also this result.  
 
The HRS experiment measured $\nc$ and $\nl$, and found $\dcl = 1.6$. 
The results on $\nc$ and $\nl$ are consistent with those found by MARK II and TPC, and 
the size of the statistical errors are similar.  
If we attribute to the HRS value of $\dcl$ a total uncertainty similar to 
those quoted by MARKII and TPC (here we assume a total error of $\pm 1.5$) 
and include also this measurement in our weighted average, we would get 
$\langle\dcl\rangle = 1.09 \pm 0.32$.  
 
In conclusion, we use $\langle\dcl\rangle = 1.0 \pm 0.4$ in the present analysis,  
and we point out that considering the current experimental precision there is no evidence  
of energy dependence of $\dcl$ in the range 29~GeV~$\leq \sqrt{s} 
\leq$~91~GeV. 
 
\subsection{About the measurement of $\dbl$} 
\label{dbl} 
 
In Table 2 we present an updated review of the experimental measurements of the 
mean charged particle multiplicities, $\nhad$, $\nb$ and $\nl$, respectively for 
the inclusive sample (when measured), $\bb$ events and $\rm \ell\bar{\ell}$ 
($\rm \ell\bar{\ell} = \light$) events. 
The difference $\dbl = \nb - \nl$ is also shown. 
The results are corrected for detector effects as well as for initial
state radiation effects.  
Charged decay products from the $K^o_S$ and $\Lambda$ decays are included.  
The quoted errors are obtained by combining the statistical and the systematic uncertainties 
in quadrature. 
The published results on $\dbl$ from OPAL, SLD, DELPHI and VENUS take correlations into account.
According to~\cite{chrin}, the DELCO result appearing in table 2 was corrected by $+25\%$ 
as compared to the published DELCO data, i.e. $3.6 \pm 1.5$, to account for the overestimated 
$b$ purity of the selected sample.

We would like to stress at this point that the results on $\dbl$ presented 
in published compilations, including this one, are not all direct  
measurements. 
MARKII and TPC at $\sqrt{s}=29$ GeV, TASSO at 35 GeV, OPAL, SLD and 
DELPHI at 91 GeV and
DELPHI and OPAL at LEP2 energies, measured $\nb$, $\nc$  
and either $\nhad$, the inclusive mean charged multiplicity, or $\nl$  
(or both), from which $\dbl$ is calculated in a direct way. 
The other experiments, instead, have only measured $\nb$ and $\nhad$,  
and, thus, one particular value for $\nc$ or $\dcl$ must be assumed in 
order to evaluate $\nl$ and $\dbl$. 
In the previous reviews, the value of $\dcl$ was the same as in 
\cite{bas}, while in the recent publication by VENUS \cite{venus} the result of 
OPAL measurement~\cite{opal1} is taken. 
In Table 2 we used for all these experiments the new average value of 
$\dcl$ presented in the previous section, $\dcl = 1.0 \pm 0.4$, and this explains  
why these results are not the same as those presented in previous publications.

\begin{table}[p] 
\begin{center} 
\begin{tabular}{|c||c|c|c|c|c|} 
\hline 
 Experiment & $\sqrt{s}$  & $\nhad$ &  $\nb$  & $\nl$   & $\dbl$            \\ 
\hline 
\hline 
DELCO~\cite{delco} & 29 & 12.3 $\pm$ 0.8 & 15.2 $\pm$ 1.3 & 11.6 $\pm$ 0.9 & 4.5 $\pm$ 1.6 \\ \cline{2-6} 
MARKII~\cite{row}  & 29 & 12.9 $\pm$ 0.6 & 16.1 $\pm$ 1.1 & 12.2 $\pm$ 1.4 & 3.9 $\pm$ 1.8 \\ \cline{2-6} 
TPC~\cite{tpc}     & 29 &                & 16.7 $\pm$ 1.0 & 12.0 $\pm$ 0.9 & 4.7 $\pm$ 1.4 \\ \cline{2-6} 
\hline 
Average & 29 & & & & 4.4 $\pm$ 0.9 \\ \cline{2-6} 
\hline 
\hline 
TASSO~\cite{tasso1} & 35 & 13.4 $\pm$ 0.66 & 16.0 $\pm$ 1.5 & 11.9 $\pm$ 1.2 & 4.1 $\pm$ 1.9 \\ \cline{2-6} 
TASSO~\cite{tasso2} & 42.1 & 14.9 $\pm$ 0.7 & 17.0 $\pm$ 2.0 & 14.3 $\pm$ 1.5 & 2.7 $\pm$ 2.5 \\ \cline{2-6} 
TOPAZ~\cite{topaz} & 58 & 14.21 $\pm$ 0.12 & 16.24 $\pm$ 1.1 & 13.57 $\pm$ 0.7 & 2.7 $\pm$ 1.3 \\ \cline{2-6} 
VENUS~\cite{venus} & 58 & 16.79 $\pm$ 0.23 & 19.38 $\pm$ 0.88 & 16.07 $\pm$ 0.7 & 3.31 $\pm$ 0.37 \\ \cline{2-6} 
\hline 
\hline 
MARKII~\cite{mark2} & 90.9 & 20.9 $\pm$ 0.5 & 23.1 $\pm$ 1.9 & 20.1 $\pm$ 0.9 & 3.0 $\pm$ 2.1 \\ \cline{2-6} 
DELPHI~\cite{delphi1} & 91.2 &    & 23.32 $\pm$ 0.51 & 20.20 $\pm$ 0.45 & 3.12 $\pm$ 0.68 \\ \cline{2-6}
OPAL~\cite{opal2}   & 91.2 &      & 23.62 $\pm$ 0.48 & 20.82 $\pm$ 0.44 & 2.79 $\pm$ 0.30 \\ \cline{2-6} 
SLD~\cite{sld2}     & 91.2 &      & 23.14 $\pm$ 0.39 & 20.21 $\pm$ 0.24 & 2.93 $\pm$ 0.33 \\ \cline{2-6} 
SLD~\cite{sld3}     & 91.2 &      & 23.098 $\pm$ 0.378 & 20.048 $\pm$ 0.316 & 3.050 $\pm$ 0.311 \\ \cline{2-6}  
\hline 
\hline 
OPAL~\cite{opal3}   & 130  &      & 25.9 $\pm$ 1.3 & 21.0 $\pm$ 1.4 & 4.9 $\pm$ 1.5 \\ \cline{2-6} 
OPAL~\cite{opal3}   & 136  &      & 25.7 $\pm$ 1.7 & 23.0 $\pm$ 1.6 & 2.8 $\pm$ 2.0 \\ \cline{2-6} 
OPAL~\cite{opal3}   & 161  &      & 24.1 $\pm$ 1.7 & 21.1 $\pm$ 2.1 & 2.9 $\pm$ 2.3 \\ \cline{2-6} 
OPAL~\cite{opal3}   & 172  &      & 28.8 $\pm$ 2.2 & 26.8 $\pm$ 2.1 & 2.1 $\pm$ 2.5 \\ \cline{2-6} 
DELPHI~\cite{delphi2} & 183 &     & 29.79 $\pm$ 1.14 & 25.25 $\pm$ 1.35 & 4.55 $\pm$ 1.5 \\ \cline{2-6} 
OPAL~\cite{opal3}   & 183  &      & 28.3 $\pm$ 1.2 & 26.8 $\pm$ 1.6 & 1.5 $\pm$ 1.6 \\ \cline{2-6} 
DELPHI~\cite{delphi2} & 189 &     & 30.53 $\pm$ 0.78 & 26.10 $\pm$ 0.97 & 4.43 $\pm$ 1.05 \\ \cline{2-6} 
OPAL~\cite{opal3}   & 189  &      & 28.89 $\pm$ 0.77 & 25.41 $\pm$ 1.0 & 3.48 $\pm$ 1.2 \\ \cline{2-6} 
OPAL~\cite{opal3}   & 192  &      & 28.5 $\pm$ 1.4 & 24.4 $\pm$ 1.9 & 4.1 $\pm$ 2.0 \\ \cline{2-6} 
OPAL~\cite{opal3}   & 196  &      & 31.3 $\pm$ 1.5 & 28.6 $\pm$ 1.6 & 2.7 $\pm$ 1.7 \\ \cline{2-6} 
DELPHI~\cite{delphi2} & 200 &     & 29.38 $\pm$ 0.82 & 25.99 $\pm$ 1.03 & 3.39 $\pm$ 1.35 \\ \cline{2-6} 
OPAL~\cite{opal3}   & 200  &      & 30.3 $\pm$ 1.3 & 25.6 $\pm$ 1.5 & 4.7 $\pm$ 1.8 \\ \cline{2-6} 
OPAL~\cite{opal3}   & 202  &      & 29.9 $\pm$ 1.7 & 25.5 $\pm$ 2.0 & 4.4 $\pm$ 2.0 \\ \cline{2-6} 
DELPHI~\cite{delprel} & 206 &     & 28.72 $\pm$ 0.77 & 24.22 $\pm$ 1.09 & 4.50 $\pm$ 1.17 \\ \cline{2-6} 
OPAL~\cite{opal3}   & 206  &      & 30.08 $\pm$ 1.0 & 26.53 $\pm$ 1.4 & 3.55 $\pm$ 1.2 \\ \cline{2-6} 
\hline 
\end{tabular} 
\end{center} 
\caption{\label{table2} 
Corrected mean charged particle multiplicities and $\dbl$ at different energies (see text
in Appendix B.2 for more details).  
According to~\cite{chrin}, the DELCO result appearing in this table was 
corrected by $+25\%$ as compared to the published DELCO data, i.e. 
$3.6 \pm 1.5$, to account for the overestimated $b$ purity of the selected sample.} 
\end{table} 
\clearpage 
\end{appendix} 

\end{document}